\documentclass[preprint2]{aastex}

\newcommand{\myemail}{maggie@physics.mcgill.ca}
\newcommand{\psr}{PSR~J1846$-$0258}

\newcommand{\rxte}{{\textit{RXTE}}}

\newcommand{\nudotdotdot}{{\ifmmode\stackrel{\bf \,...}{\textstyle \nu}\else$\stackrel{\,\...}{textstyle \nu}$\fi}}
\newcommand{\degrees}{^{\circ}}

\newcommand{\xte}{{\it RXTE}}

\usepackage{graphicx}

\shorttitle{Post-burst timing behavior of \psr}
\shortauthors{Livingstone et al.}

\begin{document}
\title{Timing Behavior of the Magnetically Active Rotation-Powered Pulsar
in the Supernova Remnant Kesteven 75}

\author{Margaret A.~Livingstone \altaffilmark{1},
Victoria M.~Kaspi}
\affil{Department of Physics, Rutherford Physics Building, 
McGill University, 3600 University Street, Montreal, Quebec,
H3A 2T8, Canada}

\and

\author{Fotis.~P.~Gavriil}
\affil{NASA Goddard Space Flight Center, Astrophysics Science Division, Code
662, Greenbelt, MD 20771}
\affil{Center for Research and Exploration in Space Science and Technology,
University of Maryland Baltimore County, 1000 Hilltop Circle, Baltimore, MD
21250}

\altaffiltext{1}{\myemail}

\clearpage

\begin{abstract}

We report a large spin-up glitch in \psr\ which coincided with
the onset of magnetar-like behavior on 2006 May 31. We 
show that the pulsar experienced an unusually large glitch recovery, with a
recovery fraction of $Q=8.7\pm 2.5$, 
resulting in a net decrease of the pulse frequency.
Such a glitch recovery has never before been observed in 
a rotation-powered pulsar, however, similar but smaller glitch
over-recovery has been recently reported in the magnetar AXP 4U~0142+61 and 
may have occurred in the SGR 1900+14. We also report a large
increase in the timing noise of the source. We discuss the implications of the
unusual timing behavior in \psr\ on its status as the first identified 
magnetically active rotation-powered pulsar.

\end{abstract}


\section{Introduction}
\label{sec:intro}

\psr\ is a young ($\sim$800\,yr), 326\,ms pulsar, discovered in 2000 with the 
\textit{Rossi X-ray Timing Explorer} \citep[\rxte; ][]{gvb+00}.
No radio pulsations have been detected despite
deep searches \citep{kmj+96,aklm08}.
\psr\ has a large inferred magnetic
field of $B\simeq5\times10^{13}$\,G, above the quantum critical limit,
and a braking index of $n=2.65\pm0.01$ \citep[][hereafter
LKGK06]{lkgk06}. The pulsar was therefore 
believed to be simply a high-magnetic field rotation-powered pulsar (RPP), with
radio pulsations that do not cross our line of sight. However,
in 2006 May, the source experienced a series of X-ray bursts and a 
sudden increase in X-ray flux, which strongly suggest magnetic activity
\citep{ggg+08}. The source also showed spectral variations and changes
in the surrounding nebula \citep{ggg+08,ks08,nsgh08}.

A magnetar is a neutron star whose magnetic energy powers the bulk
of its emission \citep{td95}, while a RPP produces
radiation via the loss of rotational kinetic energy
\citep[e.g.,][]{mt77}.
\psr\ is a unique transition object between these two source
classes. The X-ray luminosity of \psr\ can be entirely accounted
for by the spin-down power of the pulsar, however, the observed 
X-ray bursts and
flux flare are phenomena only seen thus far from magnetars
\citep{ggg+08}.

A neutron star glitch is defined as a sudden, usually unresolved,
increase in spin frequency, $\nu$. Glitches are often accompanied by an 
increase in the frequency derivative, $\dot\nu$, and are sometimes followed
by an exponential decay on timescale $\tau_d$, where some of the initial
jump in $\nu$ is recovered. In general, a glitch at time $t_g$ 
can be modeled as:
\begin{equation}
\nu(t) = \nu_0(t) + \Delta\nu_p + \Delta\nu_de^{-(t-t_g)/\tau_d} +
\Delta{\dot\nu}(t-t_g),
\end{equation}
where $\nu_0(t)$ is the frequency of the pulsar prior to the glitch
and $\Delta\nu$ is the initial frequency jump, which can be decomposed into the part
of the glitch that is permanent, $\Delta\nu_p$, and that which decays,
$\Delta\nu_d$. The recovery fraction is defined as $Q \equiv
\Delta\nu_d/\Delta\nu$.
The variety of observed pulsar
glitches is interesting. The fractional magnitude of the observed change in
$\nu$ ranges
over 6 orders of magnitude from $10^{-11}<\Delta\nu/\nu<10^{-5}$
\citep{lsg00,js06,hlj+02}. Some glitches are characterized only by a $\nu$ increase
\citep[e.g., PSR B1758$-$23;][]{sl96}, while some glitches (typically in young 
pulsars) are dominated by an increase in the magnitude of 
$\dot\nu$ \citep[e.g., the Crab pulsar;][]{wbl01}. Some glitches decay on timescales of $\sim$days
\citep[e.g.,][]{fla90}, while others display very long decay time scales \citep[$\sim$
hundreds of days; ][]{wmp+00}. 
The fraction of the glitch that decays is also highly variable. In older pulsars the
amount of glitch recovery is typically small \citep[$Q<<1$, ][]{sl96}
while in the Crab pulsar the recovery can be nearly complete
\citep[e.g.,  $Q\sim0.96$, ][]{loh81}. However, 
in many glitches, no recovery is detected.

Glitches are now known to be ubiquitous
in magnetars as well as rotation-powered pulsars
\citep{klc00,kg03,dis+03,dkg08,gdk09}. What remains
to be seen is whether glitches have the same physical origin 
in both types of objects, or if the
super-critical magnetic fields of magnetars are responsible for different
glitch origins and evolutions. While some magnetar glitches are indistinguishable from
those observed in RPPs, others occur contemporaneously with radiative
changes such as bursts, flux enhancements, and spectral or pulse profile
variations, such as in the Anomalous X-ray Pulsar (AXP) 1E~2259+586 \citep{kgw+03}. 
No radiative changes have been observed with RPP glitches \citep[e.g., ][]{hgh01}.
RPP glitches are believed to arise from a sudden
unpinning of vortices in the superfluid interior crust of the pulsar
\citep[e.g.,][]{accp93}. Magnetar glitches, on the other hand, may be
triggered by strong internal
magnetic fields as the crust is deformed, either plastically or cracked 
violently \citep{td96a}. 

In this paper, we discuss the timing behavior of \psr\ prior to,
during, and following the period of magnetic activity observed in 2006. We show that a
large glitch occurred contemporaneous with the X-ray bursts and onset of the
flux flare, accompanied by an unusual increase in the timing noise of
the pulsar. We show that the glitch recovery is very unusual for a
RPP but is reminiscent of timing behavior observed from
magnetars and is further evidence of magnetic activity in \psr. 

\section{\rxte\ Observations and Analysis}
\label{sec:obs}
Observations of \psr\ were made using the Proportional Counter Array 
\citep[PCA;][]{jsg+96,jmr+06} on board {\textit{RXTE}}. The PCA consists of an
array of five collimated xenon/methane multi-anode proportional counter
units (PCUs) operating in the 2~--~60\,keV range, with a total effective
area of approximately $\rm{6500~cm^2}$ and a field of view of 
$\rm{\sim 1^o}$~FWHM.

Our entire \rxte\ data set spans 9.7\,yr from 1999 April 18 through 2008 December 10 
(MJD 51286 -- 54810). Data from 2000 January 31 - 2005 July 27
(MJD 51574 -- 53578) were reduced and analyzed
previously and details can be found in LKGK06\nocite{lkgk06}. Eleven
observations taken in 1999 April 18-21 are of limited use for the current
analysis since they cannot be unambiguously phase connected to the rest of
the data. Analysis of data spanning 2005 July 27 - 2008 December 10 
(MJD 53578 -- 54810) is presented here. Data were collected 
in ``GoodXenon'' mode, which records the arrival time
(with 1-$\mu$s resolution) and energy (256 channel resolution) of every
unrejected event. Typically, two to three PCUs were operational
during an observation. We used the first Xenon layer of each
operational PCU and extracted events in channels 4~--~48
(approximately 2~--~20\,kev), 
as this maximizes the signal-to-noise ratio for this source.

Observations were downloaded from the HEASARC 
archive\footnote{http://heasarc.gsfc.nasa.gov/docs/archive.html}
and data from each active PCU were merged and binned at
(1/1024)\,s resolution. Photon arrival times were converted
to barycentric dynamical time (TDB) at the solar system barycenter using the
J2000 source position RA = $18^{\rm{h}}46^{\rm{m}}24\fs94\pm0\fs01$,
Decl $= -02 \degrees 58\arcmin30.1\arcsec\pm0.2\arcsec$ \citep{hcg03}
and the JPL DE200 solar system ephemeris.

\begin{figure}
\plotone{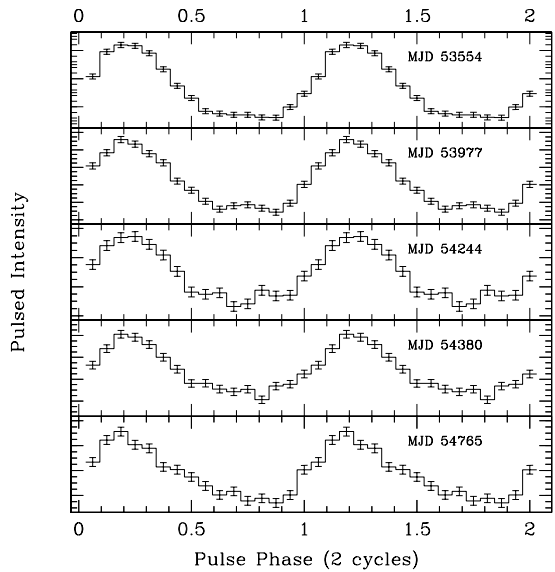}
\figcaption[fig:profiles]{\label{fig:profiles}
Five average 2~--~20 keV pulse profiles for \psr. Each profile
is based on a single phase-coherent subset of data, except
for the profile made from un-connected periodogram measurements
of $\nu$ after the burst. For each profile, the center
date of observations is listed in the top right hand corner.
The top panel shows the added
profile for a phase-coherent timing solution spanning
MJD~53228-53879,
with $\nu$, $\dot\nu$, and $\ddot\nu$ fitted. The profile
includes 29 observations with a total exposure time of 223\,ks.
The second panel shows the summed profile for 9 observations after
the X-ray bursts and flux flare spanning MJD 53949-54005 with an
exposure time of 44.3\,ks. The third panel shows the profile
for a phase-coherent timing solution of 8 observations with $\nu$
and $\dot\nu$ fitted spanning MJD 54215-54265, near the
unusual variations in $\dot\nu$. The total exposure time
is 38.9\,ks.
The fourth panel shows the average pulse profile near the end of the
glitch recovery, using a coherent timing solution of 9 observations
with $\nu$ and $\dot\nu$ fitted, spanning MJD 54363 - 54397. The
exposure
time is 40.2\,ks. The bottom profile shows the average pulse
profile after the glitch had recovered using a coherent timing
solution of 12 observations, spanning MJD 54726-54805. The exposure
time is 59.0\,ks.}
\end{figure}

The known ephemeris from LKGK06\nocite{lkgk06} was used to fold
each time series with 16 phase bins. Resulting profiles were cross-correlated 
in the Fourier domain with a high signal-to-noise ratio template created by 
adding phase-aligned profiles from all observations.
The cross-correlation process assumes that the
pulse profile is stable; indeed, we found no evidence for variability
that could bias TOA measurement, as
shown in Figure~\ref{fig:profiles}, and confirmed by \citet{kh09}. 
The figure shows average profiles from
before, after, and throughout the outburst and unusual timing behavior. 
For each individual observation, the cross-correlation yielded
the time of arrival (TOA) of phase-zero of the average pulse profile at
the fold epoch. The TOAs were fitted to a timing model (see Section \ref{sec:timing}) 
using the pulsar timing software package
TEMPO\footnote{http://www.atnf.csiro.au/research/pulsar/tempo}.
After phase-connecting the data, we
merged observations occurring on a single day and 
used the ephemeris to re-fold the data in order to obtain more
precise TOAs. This process produced 199 TOAs with a typical uncertainty
$\sim$9\,ms ($\sim$2.7\% of the pulse period). Further details of the 
observation and analysis are given in LKGK06\nocite{lkgk06}. 

\section{Timing Analysis and Results}
\label{sec:timing}
Phase-coherent timing is a powerful method for obtaining
accurate pulsar parameters, but can only be used when timing noise and
glitches are relatively small \citep[e.g.,][]{lkg05}. To obtain a
phase-coherent timing solution, each turn of the pulsar is accounted 
for by fitting TOAs with a Taylor expansion of
the pulse phase \citep{ls05a}. 

Phase-coherent timing for \psr\ spanning 2000 January 1 -- 2005 July 27
(MJD~51574--53578) is discussed in LKGK06. Phase coherence was maintained 
for the next 308 days without incident. Phase coherence was lost
with the observation occurring on 2006 May 31 (MJD~53886) which contained 4
X-ray bursts and a pulse flux increase \citep{ggg+08}. For the following 32
observations spanning 192 days, no unambiguous phase coherent timing solution
was possible. Instead, we performed periodograms to determine the pulse
frequency. Uncertainties were determined from a Monte Carlo simulation,
where noise was added to simulated sinusoidal pulses and the frequency
for each trial was determined in the same way as for the real data (see 
LKGK06 for further details). Phase coherence was once again
obtained starting 2007 January 26 (MJD~54126) with closely
spaced bootstrapping observations after the source reappeared from behind
the Sun after 48 days. A single phase coherent timing solution was obtained
spanning 2007 January 26 -- 2008 December 10 (MJD~54126--54810). This timing
solution is severely contaminated by long-term glitch recovery and timing 
noise, so the fitted parameters for the global solution are of limited value. 

\begin{figure}
\plotone{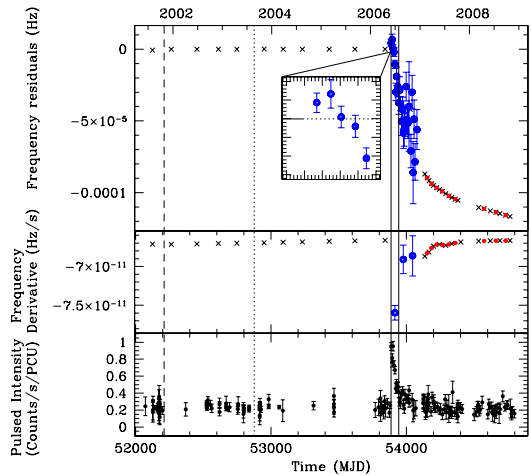}
\figcaption[fig:freq]{
\label{fig:freq}
Timing and pulsed flux evolution of \psr\ over 9
years. Top panel: Frequency measurements with pre-glitch $\nu$,
$\dot\nu$, and $\ddot\nu$ removed from all points for clarity.
Black crosses are produced from short phase-coherent measurements of
$\nu$ and $\dot\nu$, while filled red circles are produced
similarly but from overlapping segments of data. Uncertainties
are smaller than the points and are excluded for clarity. Open blue
circles are periodogram measurements of $\nu$. Middle panel: Frequency
derivative ($\dot\nu$) measurements. Crosses are produced from short
phase-coherent measurements of $\nu$ and $\dot\nu$, while filled
red circles are from overlapping segments of data, but are otherwise
produced in the same manner. See Section \ref{sec:timing} for details. 
Open blue circles are from weighted least-squares fits of periodogram $\nu$
measurements. Bottom panel: Pulsed intensity measurements in the
2~--~20\,keV energy band, as described in Section \ref{sec:flux}. The
only significant flux increase is coincident with 4 X-ray bursts
observed on MJD 53886. All panels: The dashed line represents the
epoch of a small Crab-like glitch near MJD 52210, while
the dotted line represents the epoch of a small candidate glitch
between MJDs 52837 and 52915 (LKGK06). The two solid lines show the epochs
where bursts were detected \citep{ggg+08}.}
\end{figure}

To analyze the long-term rotational behavior of the pulsar, we created
short phase-coherent timing solutions from 2000 until the onset of bursts, 
and from 2007 and 2008. Each timing solution included only $\nu$ and
$\dot\nu$, and included as much data as
possible while requiring the reduced $\chi^2$ value of the fit to be $\sim$1. This
resulted in 11 measurements of $\nu$ and $\dot\nu$ pre-glitch and 11
measurements post-glitch.
In order to better utilize the available data in the post-glitch period,
we created short overlapping timing solutions, 
each of which uses approximately half
the data from two of the above described short data subsets, and has the same
fitted parameters and $\chi^2$
requirements. This can be useful because the short coherent
timing solutions result in parameter fits that are dominated by the end points,
which can be problematic when $\nu$ is varying rapidly from glitch recovery 
as in this case, or when timing noise is a significant effect. 
This produces an additional 9 post-glitch measurements of $\nu$ and $\dot\nu$. 
Coherent frequency measurements (crosses), overlapping frequency measurements
(filled circles) and frequency measurements obtained via
periodograms (open circles), are plotted in the top panel of Figure~\ref{fig:freq}, 
with the pre-burst $\nu$, $\dot\nu$, and $\ddot\nu$ removed. 
The middle panel of Figure~\ref{fig:freq} shows measurements of $\dot\nu$ from the
short coherent timing solutions as crosses, with the overlapping $\dot\nu$ 
measurements in filled circles. In addition, open circles show three
measurements of $\dot\nu$ that are calculated from weighted least-squares fits to
nine periodogram measurements of $\nu$.

A frequency increase is apparent in the frequency residual
plot (top panel and inset, Figure~\ref{fig:freq}), indicating that a glitch occurred
\citep[also noted by][]{kh09}. Two measurements of $\nu$ are larger
than the pre-glitch predicted value. We calculated the average
$\Delta\nu = (5.5\pm 2.3) \times 10^{-6}$ at MJD~53890 (4~--~11\,days
after the glitch occurred). The initial value of $\Delta\nu$ at the
time of the glitch was presumably larger than this value, as indicated by
the exponential fit to the data discussed below. 

\begin{figure}
\plotone{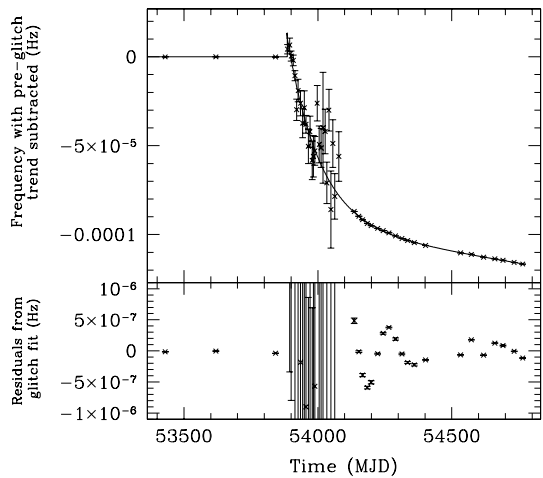}
\figcaption[freq]{Top panel: Frequency measurements of \psr\ (points)
and the fitted exponential recovery glitch model (solid line).
Pre-glitch $\nu$, $\dot\nu$, and $\ddot\nu$ have been removed from
all frequency measurements. Bottom panel: Residuals from the fit
with the formal, un-multiplied uncertainties. The best fit has a
${\chi^2}_\nu$ of $\sim 267$ for 45 degrees of freedom. Significant
variation from the fitted exponential is clear, however, the
exponential recovery dominates the change in $\nu$ by $\sim2$ orders
of magnitude over the remaining systematic variations in $\nu$.
Given the large $\chi^2$, we increased the uncertainties on the
phase-coherent $\nu$ measurements until ${\chi^2}_\nu \sim 1$,
and report uncertainties on the fitted model from the $\Delta\chi^2=1$
contours. All fitted parameters are given in Table~\ref{table:glitch}.
\label{fig:glitchresids}}
\end{figure}

We fitted the measured frequencies of \psr\ (fitting only 3 pre-glitch
$\nu$ values) with an exponential recovery
glitch model (Eq. 1), the results of which are shown in
Figure~\ref{fig:glitchresids}. The top panel of the Figure shows 
frequency measurements with the pre-glitch ephemeris ($\nu$, 
$\dot\nu$ and $\ddot\nu$) removed
(as in Figure~\ref{fig:freq}), while the bottom panel shows
residuals from the glitch fit. The uncertainties from the periodogram
measurements of $\nu$ are $\sim2-3$ orders of magnitude larger than those
from short coherent fits to the data and thus contribute minimally to the
overall $\chi^2$ value. Thus the glitch residuals (bottom panel,
Figure~\ref{fig:glitchresids}) are shown to highlight the
deviation from the fit of the coherent frequencies, which dominate the 
$\chi^2$. 

Significant deviations from the fit can be seen during the period of
glitch recovery, giving rise to a large reduced $\chi^2$ value of
$\sim267$ for 45 degrees of freedom for the best fit (bottom panel,
Figure~\ref{fig:glitchresids}).

Given the poor fit, the formal uncertainties underestimate the true
values. To determine more reasonable uncertainties on the glitch
parameters, we multiplied the
uncertainties on the coherent $\nu$ measurements
by a factor until the reduced $\chi^2$ of the fit $\sim 1$. We
applied a multiplicative factor only to the phase coherent $\nu$
measurements because these uncertainties were $\sim 2-3$ orders of
magnitude smaller than those from periodogram $\nu$ measurements,
determined in a different way, and the contribution to $\chi^2$
was mainly from the coherent $\nu$ values. Quoted uncertainties on
the glitch parameters are from $\Delta \chi^2=1$ contours from the
fit with multiplied uncertainties. The fitted value of the
initial fractional frequency increase is $\Delta\nu/\nu=4.0(1.3)\times10^{-6}$,
very large for such a young pulsar, where a typical value is
$\Delta\nu/\nu \sim 10^{-8}$. More remarkable yet, however, is 
the amount by which the frequency recovers. We find $Q=8.7(2.5)$,
corresponding to a net decrease in frequency of $\Delta\nu_p =
9.52(9)\times10^{-5}$\,Hz. All fitted glitch parameters are 
given in Table~\ref{table:glitch}.

While the deviation from the exponential fit is very significant for 
several months (bottom panel, Figure~\ref{fig:glitchresids}) the overall evolution
after the glitch is dominated by the exponential recovery: the deviation
from the fit is $\sim$2 orders of magnitude smaller than the overall
post-glitch decrease in $\nu$.
The deviation from exponential recovery may well have been larger in the
period just following the glitch, however, the large uncertainties on the
periodogram measurements of $\nu$ ($\sim10^{-6}$\,Hz)
prevent any firm conclusion. However, since the corresponding pulse TOAs 
cannot be unambiguously phase-connected, 
it is likely that large variations in $\nu$ and $\dot\nu$ occurred
during the 240-day period between the glitch epoch and 
when we regained phase-coherence.
It is also possible that a second, smaller glitch 
($\Delta\nu/\nu < 10^{-7}$) occurred during this period. 
The observed deviation from the exponential recovery
decreases as the glitch recovers. Thus, in the
closing months of 2008, the pulsar was rotating very regularly again, similar
to its pre-glitch behavior. 

\begin{figure}
\plotone{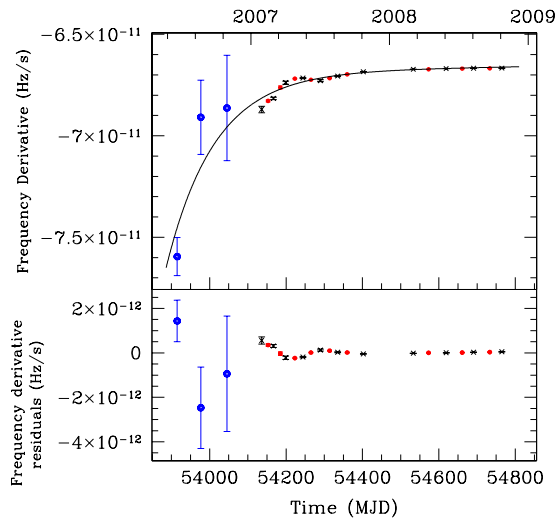}
\figcaption[freq]{Top panel: Post-burst $\dot\nu$ measurements of \psr.
Black crosses are produced from short
phase-coherent measurements of $\nu$ and $\dot\nu$, while filled red 
circles are from overlapping segments of data, but are otherwise produced
in the same manner. Open blue circles are from weighted least-squares
fit to periodogram measurements of $\nu$. See Section~\ref{sec:timing}
for details. Note the significant deviation from a simple exponential
glitch recovery evident in the coherent $\dot\nu$ measurements in the
interval MJD~54100--54300. The solid line is the derivative of the
exponential recovery model fitted to the frequency data. Bottom
panel: Residuals of the $\dot\nu$ measurements and derivative
of the exponential glitch recovery model.
\label{fig:nudot}}
\end{figure}

Figure~\ref{fig:nudot} shows post-glitch phase-coherent measurements of $\dot\nu$ 
(a subset of the $\dot\nu$ measurements shown in the middle panel of
Figure~\ref{fig:freq}). The pre-glitch measurements
are excluded here for clarity. The solid line is the derivative of the
glitch model fitted to the $\nu$ measurements, 
clearly showing that there is significant deviation.
The overall effect of the glitch recovery on
$\dot\nu$ is clear, however, from MJD 54100-54300 the $\dot\nu$
measurements deviate from the exponential recovery by $\sim 0.15$\%. 
The effect of this anomalous change in $\dot\nu$ 
is not directly evident in the measurements of $\nu$ (which
are dominated by the exponential recovery) but does help explain why the
exponential glitch fit is not a satisfactory description of the data,
and is clear in the residuals of the glitch fit in the bottom panel 
of Figure~\ref{fig:freq}. 

\section{Bursts and pulsed flux} 
\label{sec:flux}

The glitch during PSR~J1846$-$0258's outburst was accompanied by a
major pulsed flux enhancement \citep{ggg+08}.  In order to quantify
the radiative properties of the source during the glitch recovery, we
extracted its pulsed flux using all available \xte\ observations.
First, we generated separate event lists for each PCU in
FITS\footnote{\url{http://fits.gsfc.nasa.gov}} format using the
standard \texttt{FTOOLS}\footnote{\url{http://heasarc.gsfc.nasa.gov/docs/software/ftools/}}.
We then filtered our event lists such that we only preserved photons
in the 2~--~20\,keV band, from the first Xenon layer.  The photon
arrival times were then barycented using the source position and the
JPL DE200 solar system ephemeris.  We folded the filtered barycented
photon arrival times using the ephemeris determined in our
phase-coherent timing analysis using 16 phase bins. Using the folded
profiles, we calculated the RMS pulsed flux in each PCU using the
Fourier method described by \citet{wkt+04} keeping only the
contribution from the 1st harmonic given the source's roughly sinusoidal
profile. Not all the observations were pointed at PSR~J1846$-$0258,
therefore we corrected for the reduced efficiency in each PCU due to
the offset pointing using the collimator response of each PCU and the
instrument attitude files.  Finally, we averaged the pulsed flux
in each PCU weighted by the fractional exposure of each PCU.  
We excluded the contribution from PCU 0 because of the loss of its
propane layer and the numerous detector breakdown events.
Our pulsed flux time series is presented in the bottom panel of
Figure~\ref{fig:freq}. The event lists for each PCU created for the 
pulsed flux analysis were
binned into 31.25~ms lightcurves and were searched for bursts using
the burst search algorithm introduced in \citet{gkw02}.  No additional
bursts were found other than the 5 reported in \citet{ggg+08}.

\begin{figure}
\plotone{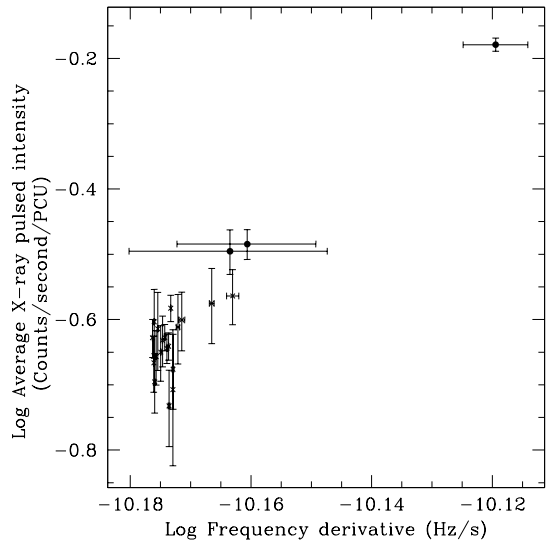}
\figcaption[torqueflux]{Observed relationship between pulsed flux 
and the spin-down rate of \psr. The measurements of $\dot\nu$ from
phase-coherent timing solutions before and after the outburst are
shown as crosses. The three $\dot\nu$ measurements made from least
squares fits to periodogram measurements of $\nu$ are shown as 
filled circles. The best-fit power-law index to all data is $\sim 8$,
however, the correlation is constrained by a single measurement of
$\dot\nu$, made when the timing noise was extremely large. Excluding
this single point results in no significant correlation.
See Section~\ref{sec:flux} for details.
\label{fig:torqueflux} }
\end{figure}

We checked for a correlation between torque and pulsed flux by plotting the
pulsed flux against the spin frequency derivative in log-log space, shown in
Figure~\ref{fig:torqueflux}. All measurements of $\dot\nu$ before
and after the glitch and bursts are shown, 
including 3 measurements of $\dot\nu$ obtained from weighted 
least-squares fits to $\nu$ measurements obtained from periodograms.
These 3 $\dot\nu$ measurements have uncertainties $\sim2-3$ orders of
magnitude larger than those for $\dot\nu$ measurements from coherent 
timing. 

The plot shows a possible correlation between flux and torque when
both parameters are extreme. Fitting a power law to the data gives a
power law index of $\sim$8. However, this 
value is constrained by a single measurement of $\dot\nu$ when
the flux, and timing noise, are both very large. Excluding this point, 
there is no significant correlation between flux and $\dot\nu$. 
The observed variations in $\dot\nu$ during the later phase connected
period, are at the $\sim3\%$ level, while the variation in
$\dot\nu$ immediately following the glitch are at the $\sim 15\%$
level (Figure~\ref{fig:nudot}). Correspondingly small fluctuations 
in pulsed flux are not detectable in these data.

\section{Discussion}
\label{sec:discussion}

\subsection{Glitch properties}

Four X-ray bursts in \psr\ coincided with the onset of a flux flare on 2006
May 31 (MJD 53886). The
pulsed flux decayed over $\sim$2 months and reached quiescence around the time of
the fifth burst on 2006 July 27 (MJD 53943) \citep{ggg+08}.
Significant spectral changes also occurred 
\citep{ggg+08,ks08,kh09} and flux enhancement up to 300\,keV was observed
\citep{kh09}. Contemporaneous with the sudden change in the X-ray 
emission from \psr, we
observed a large glitch with an initial frequency increase of
$\nu = 1.24(41)\times10^{-5}$.  The glitch decayed over 
$127(5)$\,days, with a recovery fraction of $Q=8.7(2.5)$,
resulting in a net decrease of the pulse frequency of 
$\Delta\nu = -9.52(9)\times10^{-5}$\,Hz. Furthermore,
the timing behavior during the period of recovery is not well
modeled by a simple exponential function and measurements of
$\dot\nu$ in particular are suggestive of a high level of 
timing noise for several hundred days following the glitch. 

This glitch and subsequent recovery reinforces that \psr\ underwent a period
of magnetic activity in 2006. This glitch is entirely different from the
previous glitch in this source, which was radiatively silent, small in magnitude
($\Delta\nu/\nu = 2.5(2)\times 10^{-9}$), dominated by a change in
$\dot\nu$ ($\Delta\dot\nu/\dot\nu =9.3(1)\times10^{-4}$), and had no
measurable recovery (LKGK06)\nocite{lkgk06}. This small glitch is 
similar to those observed in the Crab pulsar \citep[e.g.,][]{wbl01} and other
very young rotation-powered pulsars such as PSR~B0540$-$69 \citep{lkg05}. It is 
unusual to have two such disparate
initial $\Delta\nu$ magnitudes in the same source, particularly in such a young
pulsar, though this has been seen in some older pulsars such as PSR~B1737$-$30,
which has glitch magnitudes spanning four orders of magnitude
\citep{lsg00,js06}. In fact, the glitch reported here is the largest glitch ever observed in any
of the pulsars with characteristic ages less than $\sim 2$\,kyr (the Crab
pulsar, B0540$-$69, B1509$-$59, and J1119$-$6127), none of which have
experienced glitches with fractional magnitudes larger than 
$\Delta\nu/\nu \sim 10^{-8}$. 

The glitch reported here has a recovery fraction of 
$Q = 8.7\pm 2.5$. The measured value of 
$Q>1$ implies that the net frequency change after the glitch recovery
is negative, $\Delta\nu_p = -9.52(9)\times10^{-5}$, 
as shown in Figure~\ref{fig:freq}. A similar effect
was recently observed in the AXP 4U 0142+61, though with much smaller
magnitude of $Q=1.07\pm0.02$ \citep{gdk09}. The 
negative change in $\nu$ resulting from the
over-recovery of the \psr\ glitch is similar in magnitude 
to the unresolved timing event seen in the magnetar SGR 1900+14
in the months before the giant flare in
1998 \citep{wkp+99}. In that case, an enhanced spin-down of
the magnetar was observed over $\sim$3 months. Well spaced 
timing observations around the time
of the timing event were not available, so no glitch could be resolved, if
indeed one occurred. 

\citet{tdw+00} attribute the observed behavior in SGR 1900+14 
to one of two 
possibilities. The first is an increase in the magnitude of $\dot\nu$ 
by a factor of $\sim2.3$, 
persisting for $\sim80$\,days. The second is that a negative glitch,
that is, a sudden spin-down occurred, with magnitude
$\Delta\nu/\nu \simeq 1\times10^{-4}$.
However, a timing event similar to that observed in \psr\ 
could also describe the data. It is curious, however, that such 
similar fractional changes in $\nu$ should
occur in two sources that experienced such disparate radiative changes, with
the energy output from SGR~1900+14 being several orders of magnitude larger than
from \psr. 

AXP 1E~2259+586 experienced a glitch contemporaneous with 80 X-ray
bursts, a flux flare and pulse profile changes in 2002
\citep{kgw+03,wkt+04}. Neither this glitch nor the large \psr\ glitch
can be described with a single exponential decay model. 
The addition of an exponential growth component
better describes the 1E 2259+586 glitch, however, the addition
of a similar component for the \psr\ glitch does not provide a 
significant improvement to the fit.
Interestingly, the flux enhancement observed in 1E~2259+586
lasted much longer \citep[$>2$\,yr; ][]{zkd+08}  than the glitch recovery
time scale ($\tau_d \sim16$\,days), whereas the reverse
is true for \psr, with $\tau_d=127$\,days and a flux decay timescale of 
$55.5\pm5.7$\,days \citep{ggg+08}.
The 1E~2259+586 event can also be distinguished from the \psr\ glitch in that 
its recovery fraction is much smaller, with $Q\simeq 0.19$. In 2001, the AXP 
1RXS~J170849.0$-$400910 also experienced a glitch with recovery that was not
well described by
a simple exponential, and not improved with the addition of a second
exponential term \citep{kg03,dis+03,dkg08}. \citet{wkt+04} argued that
it is unlikely that 1RXS~J170849.0$-$400910 experienced bursts or a pulsed
flux flare associated with this glitch because a flux flare would have
had to decay on a time scale less than the glitch decay time scale.
However, this is exactly the behavior observed from \psr, albeit with 
a much longer time scale. Long-term spectral changes and flux
variations have been claimed in 1RXS~J170849.0$-$400910
\citep{roz+05,cri+07,igz+07}. 

An interesting characteristic of some AXP glitches is a period of enhanced
spin-down immediately following the glitch, as observed in 1E~1841$-$045,
1RXS~J170849.0$-$400910, and 
1E~2259+586 \citep{kg03,dkg08}. The additional instantaneous spin-down at the 
glitch epoch owing to the exponential recovery 
can be quantified as $\dot\nu_{\rm{inst}}=- \Delta\nu_d/\tau_d$
(found by taking the derivative of the exponential term 
in Eq. 1 and setting $t=t_g$). For the
1E~2259+586 glitch, $\dot\nu_{\rm{inst}}=(8.2\pm0.6)\dot\nu$,
while a typical value for a RPP is
$\sim 0.005\dot\nu$ \citep[see ][ and references contained therein]{per06}. 
For \psr\ the instantaneous spin-down
is $\dot\nu_{\rm{inst}}=0.15\dot\nu$, larger than for any RPP glitch, but not as 
large as those measured for AXPs. It should be noted, however, that an
enhanced spin-down is not observed in every AXP glitch. 

In \psr, there is evidence for a $\sim$200-day interval where $\dot\nu$ 
deviates significantly from an exponential
glitch recovery, and it is possible that further significant
deviation occurred during the 240-day period
of unconnected data directly following the glitch.
Perhaps $\dot\nu$ is varying in a stochastic fashion similar to
that observed in the AXP 1E 1048.1$-$5937
\citep{gk04,dkg09}. In this AXP, a $\sim$year-long period of rapid
$\dot\nu$ variations followed a large pulsed flux flare and a possible
glitch in 2002. Another glitch in 2007 coincided with the onset of a
pulsed-flux flare, again followed by stochastic variations in $\dot\nu$.
Another similarity between these sources is that the flux enhancement 
decayed away long before the timing variations subsided.
Alternatively, perhaps the variations are more simply attributed to 
timing noise, as is seen 
in many young RPPs. The behavior remains unusual however, since this is
qualitatively very different from the mild timing noise observed in \psr\
prior to magnetic activity, and such a large
change in timing noise behavior is unprecedented among RPPs. 

A simple estimate of the transfer of rotational kinetic energy
at the time of the glitch can be obtained by treating the star as a solid
rotating body. Assuming the canonical neutron star moment of inertia of
$I=10^{45}$\,g\,cm$^2$, the energy deposited in the solid crust at the
time of the glitch owing to the increase in $\nu$ is
$\Delta E \simeq (2 \pi)^2 I \nu \Delta\nu \sim 2\times10^{42}$\,ergs.
However, it is well established that neutron stars are comprised of at 
least two components, a solid crust coupled to the core 
and a loosely coupled superfluid component in the crust. Taking the
two-component model into account but not making any further
assumptions about the nature of the two components, 
we can calculate the energy transferred between the components 
at the time of the glitch. The first constraint is that angular
momentum is conserved, that is, 
\begin{equation}
I_{C} \Delta\nu = I_{SF} \Delta\nu_{SF},
\end{equation}
where $\Delta\nu$ is the observed increase in spin-frequency, and
$\Delta\nu_{SF}$ is the unknown change in spin-frequency of the superfluid,
and $I_C$ and $I_{SF}$ are the moment of inertia of the solid crust and
core, and superfluid, respectively. An estimate of the 
energy contained in the glitch can then be calculated as
\begin{equation}
E_g = \Delta E_C - \Delta E_{SF} = (2 \pi)^2 I_C \Delta\nu (\nu - \nu_{SF}).
\end{equation}
The frequency lag between the crust and superfluid
can be estimated from the glitch decay time, $\tau_d$, as
\begin{equation}
\nu_{SF} - \nu\ \simeq \frac{\tau_d \nu}{\tau_c},
\end{equation}
where $\tau_c$ is the characteristic age of the pulsar \citep{st83}. For
\psr, the frequency lag is $\nu_{SF} - \nu\ \sim 0.0013$\,Hz, 
giving a glitch energy of $\sim7 \times 10^{39}$\,ergs for the
measured glitch parameters (see Table~\ref{table:glitch}). If the
measured glitch decay timescale, $\tau_d$, in this case 
is dominated by processes external to the neutron star (as discussed 
in Section~\ref{sec:magglitchdisc}), any other glitch
decay timescale would occur on a shorter timescale than the one
observed, so this estimate of the frequency lag and thus 
glitch energy can be considered an upper limit. 

For a distance to the pulsar of 6\,kpc \citep{lt08} the energy estimated
to have been released in the bursts and flux flare, assuming isotropic
emission, is
$(3.8-4.8)\times10^{41}(d/6{\rm{kpc}})^2$\,ergs (2~--~60\,keV).
A new estimate of the distance to the pulsar of $\sim$10\,kpc
\citep{scy+09} increases the amount of energy contained in the
radiative outburst to 
$(1.1-1.3)\times10^{42}(d/10{\rm{kpc}})^2$\,ergs (2~--~60\,keV).
For either distance, the energy contained in the radiative 
outburst is several orders of magnitude larger than 
that contained in the glitch. This is similar to the bursts
and glitch from 1E~2259+586, for which the energy contained in the
glitch was $\sim$2 orders of magnitude less than the energy contained
in the bursts and flux flare \citep{wkt+04}. This is suggestive that
the glitch alone is not responsible for the radiative outburst, in
contrast to the argument put forward by \citet{kh09}. Moreover, there
is no evidence for this event (or in any other similar AXP event) that
the glitch preceded the radiative event. This could be tested only
by sensitive continuous X-ray monitoring of this and other similar
sources.

\subsection{Physical models for ``magnetic glitches''}
\label{sec:magglitchdisc}
Rotation-powered pulsar glitches are thought to arise from differential
rotation in the neutron star, where the crust contains superfluid neutrons
rotating more rapidly than the surrounding matter \citep[e.g., ][]{aaps84a,ap93}.
The angular momentum of a rotating superfluid is quantized in vortices which
are thought to
become pinned to nuclei in the star's crust. The vortex lines are therefore
under extreme stresses due to the differential rotation between the crust
and the superfluid. It is thought that the vortex lines experience
sudden unpinning, resulting in the transfer of angular momentum to the crust,
observed as an increase in the spin-frequency of the neutron star. 
Magnetar glitches may instead be triggered by strong internal magnetic fields as the
crust is deformed, either plastically or cracked violently \citep{td96a}.
This idea is supported by the large number of glitches now observed to
occur at the same epoch as magnetically powered radiative events, such as bursts and
flares \citep[e.g., ][]{kgw+03,dkg09,icd+07}. 

The physics underlying glitches with $Q>1$ is unclear. The classical
glitch model of vortex unpinning in the superfluid crust of the neutron star 
does not readily produce such dramatic glitch recoveries. 
One possibility is that some parts of the superfluid are in fact rotating
more slowly than the crust. Then the initial $\nu$ increase would be from a
transfer of angular momentum from a more rapidly rotating region of the
superfluid to the crust, which is then followed by a transfer of angular momentum from
the crust to the more sluggish region of the superfluid. This is 
the explanation put forward by \citet{tdw+00} to explain the net spin-down
event in SGR 1900+14. They argue that regions of slowly rotating superfluid
can occur in magnetars because vortex motion is dominated by advection
across the neutron star surface by the deforming crust and that gradual
plastic deformation of the neutron star crust will cause the superfluid to
rotate more slowly than the crust. However, it is not clear whether such
behavior is expected in a neutron star with a magnetic field of
$\sim5\times10^{13}$\,G, spinning relatively rapidly compared to the magnetars.

That the recovery of the \psr\ glitch so far overshoots the initial
frequency increase is suggestive of an external torque following the glitch. 
Previous evidence for an external torque includes the large
fraction of $I$ implied to have decoupled in the
1E~2259+586 glitch. However, this may also be explained with a core glitch, that is,
where the core of the pulsar (and thus a large fraction of $I$) 
decouples temporarily from the crust \citep{kg03,wkt+04}.
In \psr, however, the post-glitch relaxation amplitude is much greater than
the initial glitch amplitude, offering support for the idea 
that the post-glitch spin-down behavior results from an external source. 

One possibility is that a magnetic field twist responsible for the
X-ray bursts and flux enhancement also affects the spin-down of the
pulsar. In this case the observed recovery is driven by the propagation
of magnetic field untwisting (similar to a 
shock wave) through the magnetosphere. During this process, the spin-down of
the star may increase because the effective magnetic field has increased. 
When the ``shock'' reaches the light 
cylinder, which can take place on few month time scales, 
the spin-down should return to its pre-burst value \citep{bel08}. This
theory also allows for non-monotonic behavior in the spin-down after an
event, as observed in both AXP 1E~1048.1$-$5937 \citep{gk04,dkg09} and 
\psr. This model allows
for a delay between flux variations and the onset of timing variability, as 
was observed in 1E~1048.1$-$5937, though not observed in \psr. However, as
in the model of variably rotating superfluid, it is not immediately clear
how the more rapid rotation and smaller magnetic field affect the
relevance of this model. In particular, the different light cylinder
radii of \psr\ compared to the AXPs should result in different shock
propagation times. 

Alternatively, it has been proposed that fallback disks from the supernova
explosion creating the neutron star could be interacting with magnetars and
be responsible for some of the observed emission \citep[e.g., ][]{chn00,alp01}.
In this case, the initial X-ray bursts could irradiate a fossil disk
commencing a period of disk activity. The interaction between the neutron 
star and a disk could cause the enhanced spin-down, which would decay as the disk cooled.
However, in the framework of this model, it is difficult to understand how accretion 
causing variations in the spin-down rate could continue for so much longer than the
pulsed-flux enhancement. 

\subsection{Magnetar and High-B Radio Pulsar Properties}

Another RPP, PSR~J1119$-$6127, has very similar spin properties
to \psr. Its $P$, $\dot E$, and $\tau_c$ are all similar, and notably, it has a
similarly large magnetic field of $B=4.1\times10^{13}$\,G \citep{ckl+00}. This 
pulsar 
has shown some indication of unusual X-ray emission. No
magnetospheric X-ray emission is detected, but
thermal pulsations with a $\sim75\%$ pulsed fraction and a large surface
temperature are detected \citep{gkc+05}. No direct evidence of
magnetic activity (i.e. bursts or flux enhancements) is present in this
pulsar. Given the similarities between PSR~J1119$-$6127 and \psr, both
sources, as well as other high B-field RPPs are currently 
being monitored with \xte\ for similar magnetic activity.

\psr\ may be related to the transient AXPs (TAXPs). \psr\
appears to be a typical RPP for $\ge 95\%$ of the time, with 
brief periods of magnetic activity occurring approximately once a
decade. The TAXP XTE~J1810$-$197 increased in
brightness by a factor of $\sim$100 and was subsequently visible for 
several years as a 
pulsed X-ray source \citep{ims+04,hg05}. Another TAXP,
1E~1547$-$5408 was detected as a pulsed radio source after an X-ray outburst
in which the flux increased by at least a factor of 16 \citep{crhr07}.
Interestingly, both XTE~J1810$-$197 and 1E~1547$-$5408, two 
\textit{bona fide} TAXPs, are the only two magnetars with
detected radio pulsations. By contrast, no radio pulsations have been 
detected from \psr\
despite extensive searches both before and after the magnetic activity
\citep{kmj+96, aklm08}. 

\section{Conclusions}
In \xte\ observations of \psr, 
we have observed a large glitch with an unusual quasi-exponential over-recovery of
$\nu$ and substantial timing noise contemporaneous with X-ray bursts and a flux 
increase. These
observations strengthen the tie between magnetic activity in neutron stars and
unusual glitch activity, as has been previously noted \citep[e.g., ][]{dkg08}.
A glitch with recovery fraction $Q>1$ has never before been observed from
a rotation-powered pulsar and is not compatible with the standard model of
pulsar glitches.  The unusually large glitch recovery reported here for 
\psr, as well as the radiative changes occurring contemporaneously
with the \psr\ glitch and several AXP glitches, together provide the best
evidence that there are physical differences between typical RPP glitches
and some glitches observed in magnetars. 

Ongoing timing observations of 
\psr\ are required to obtain a deterministic
post-outburst braking index measurement that is uncontaminated by long-term
glitch recovery. A measurement of the braking index for this source after 
outburst is of
considerable interest as a change would strongly suggest that a magnetic
reconfiguration occurred at the time of the outburst.

\acknowledgments
We thank an anonymous referee for helpful comments that greatly
improved the text. We thank A. Beloborodov, D. Eichler, 
E.V. Gotthelf, and B. Link for useful discussions relating to the
text. This research made use of data obtained from the High Energy Astrophysics
Science Archive Research Center Online Service, provided by the NASA-Goddard
Space Flight Center. MAL is an National Science and Engineering Research
Council (NSERC) PGS-D fellow. VMK holds the Lorne Trottier Chair in
Astrophysics and Cosmology and a Canada Research
Chair in Observational Astrophysics. Funding for this work was 
provided by NSERC Discovery Grant Rgpin 
228738-03, FQRNT, CIFAR and CFI.

\clearpage

\begin{center}
\begin{deluxetable}{lc}
\tablecaption{Glitch Parameters for \psr. \label{table:glitch}}
\tablewidth{0pt}
\startdata
Parameter & Value \\
\hline 
\hline
Epoch (Modified Julian Day)            & 53883.0(3.0)\\ 
$\Delta{\nu}/\nu$                      & $4.0(1.3)\times10^{-6}$\\ 
$\Delta{\dot{\nu}}/{\dot{\nu}}$        & 0.0041(2)\\ 
$\tau_d$ (days)                        & 127(5)\\ 
$Q$                                    & 8.7(2.5)\\ 
\hline
${\Delta{\nu}_p}$            & $-9.52(9)\times10^{-5}$ \\ 
${\Delta{\nu}_d}$            & $10.8(4)\times10^{-5}$  \\ 
\hline
\enddata
\\
{Figures in parentheses are uncertainties in the last digits quoted
and are the estimated $1 \sigma$ uncertainties from the fitted
exponential glitch recovery model using $\Delta \chi^2=1$ contours.
The model is fitted to data with uncertainties that are increased by a
factor until ${\chi^2}_\nu \sim 1$ (see Section \ref{sec:timing} for
details).}
\end{deluxetable}
\end{center}

\end{document}